# Ultra-compact Highly Directional Pixel Technology

AMIR DJALALIAN-ASSL[1,*]

[1]*51 Golf View Drive, Craigieburn, VIC 3064, Australia*

*amir.djalalian@gmail.com

**Abstract:** Attributes such as the Radiative Decay Rate (RDR) and the Radiation Pattern (RP) of Quantum Dots (QDs) are highly sensitive to their nearby surrounding material. To enhance the RDR and shape the RP, interspacing between the QDs and their distances from discontinuities caused by multi-layered environment in which they are immersed, must be controlled with nanoscale accuracy. State of the art in QD based micro-display has so far ignored these aspects of light-matter interaction in pixel design and therefore has been unable to harness the full potential of QDs in pixel size reduction. I propose a novel pixel technology for dynamic micro-displays with novel capabilities such as (1) Each submicron pixel being operable electrically yet independently (i.e. the ability of turning a pixel on/off being independent from the neighboring pixels, hence the infinite contrast ratio. (2) Highly directional light being emitted from each pixel and (3) Increased quantum yield and luminescence while decreasing the number of QDs per pixel, hence minimizing the power consumption.

## 1. Introduction

Visual communication has been a vital part of our day to day interaction ever since history recalls. Today's technology has made it possible for almost everyone in the modern world to have access to mobile devices that can perform a variety of functions, from being a simple phone to a GPS finding directions. The visual part of conveying information is becoming more prevalent. From a TV in a living room to the computerized refrigerator in the kitchen, to a GPS device in the car, all rely on visual communication, and in that they share a fundamental technology, that is a display that comprises of "pixels", that is the smallest element in a display device that can be turned on or off independently. There are so many benefits in miniaturization of digital displays. Imagine a world where there are no street signs, and yet one is able to "see" the street name and find directions. Imagine driving in a car with a dashboard that has no monitors such fuel gauge or speedometer, and yet the driver could see all the vital information right in front of him as well as being guided by the GPS towards the destination without taking eyes off the road. Imagine an intern performing a crucial surgery in an underprivileged hospital and receiving visual instructions, (from experts far away), without taking eyes off the patient. All these may be achieved via "augmented or virtual reality" by which the images our eyes receive from a real object are overlaid by artificial images produced by a digital display. Virtual/augmented reality devices have ample of applications, from gaming industries to military to medical surgeries. But central to this is a digital display so small that could be integrated with an eyewear so our body is free to move and our hands are free to do what needs to be done. Recent advances in science and engineering has focused on miniaturization of digital displays. Research and development in the area is gaining momentum. Industries are investing more and more in finding better ways of reducing the size of the display devices while increasing their resolution. But this comes at a cost. So far, the existing pixel technology has reached its limits when one considers the trade-off between the size reduction and resolution as well as production cost.

Commercially available solutions such as Liquid Crystal Displays (LCD) are transmissive and rely on Light Emitting Diodes (LED) as backlight, hence the lateral dimension of a pixel is dictated by the backlight LED. Furthermore, the color gamut in LCDs is limited to the emitted spectrum of the backlight LED and the filtering quality of the LCD. For a while, plasmonic color filters[1, 2] as small as few hundreds of nanometers [3], were thought to be a better candidate for the new pixel technology as they possess a superior filtering quality of the white backlight. But the limiting factor remains the same, i.e. the LED based backlight technology. On the other hand, though the displays based on Organic Light Emitting Diodes (OLED) are emissive (hence no need for the backlight), their gamut is inferior to that of QDs. Furthermore, neither LCD nor OLED allow for a pixel to be fully in an "off" state. It is therefore intuitive to take advantage of inherent properties of QDs, such as high luminance efficiency, photo- and thermal- stability, cost-effectiveness, energy-efficiency, wide-range of colors, their ease of fabrications and integration in a novel pixel technology. There are reports on the collective excitation of ensembles of trapped QDs, fabricated by randomly dispensing the QDs over prepatterned metallic *meta*-surfaces consisting of arrays of rectangular nanocavities (excited with backlight not electrically), yet with no boundaries that may constitute pixels [4]. Furthermore, it is well known that the emission of QDs are quenched upon contact with metallic surfaces. Recent advances in electrohydrodynamic jet printing have led to patterning of flat surfaces with QD ensembles with lateral dimensions spanning only a few hundred nanometers [5]. Although such technologies are somewhat accurate in positioning the bundles of QDs laterally, one must also consider the cost/benefit aspect of such fabrication technique with respect to the production line. Jet printing of QDs is both slow and expensive. An alternative fabrication technique is intaglio transfer printing [6], (i.e. a variation of the microcontact printing) where the image to be displayed (though pixilated) is statically set during the fabrication, hence more appropriate for billboard-like applications rather than dynamic contents. In summary, state of the art in QD based pixel technology is facing serious limitations in:

**(a)** Reduction of pixel size while preventing inter-pixel cross contamination which leads to cross-talk-like effects.

**(b)** Provision of electrical interconnects for discretization of pixels, allowing their on/off state to be independent of the state of other pixels.

One can trace the root cause of these limitations to a common factor, that is the lack of control (and thereby a technique) in positioning the QDs in their designated location. Therefore, gaining control over QDs distribution would solve these problems. My preliminary studies have shown that, counterintuitively, stacking QDs along the optical axis is one of the least efficient ways to obtain an enhanced emission. Consequently, spreading the QDs laterally over the target area of a color element, is the most viable approach. In effect, one must find a compromise between the quantum yield and the pixel size. The reduction in pixel size and the consequent reduction in the number of QDs per pixel must be compensated by other means to maximize AND focus their emission along the optical axis, preventing the emission from spreading laterally inside the multi-layered media. But sadly, these aspects of research are non-existent in the state of the art:

**(c)** Enrichment of QD based pixel emission by means of optimization and coupling of various physical phenomena that enhance the radiative decay rate (RDR) of QDs to minimize power conception while maximizing quantum yield.

**(d)** To control the radiation pattern (RP) of a pixel by producing a unidirectional emission along the optical axis to prevent wasteful lateral scattering.

One factor hindering the progress in these areas is, again, accurate positioning of QDs within a multi-layered structure (e.g. the spacing between the QDs and the nearby metallic/dielectric interface). In fact, the lack of attention and research in these areas, often leads to suppressed emission in QD based pixels.

So, the critical technical issue that remains to be solved is how to accurately bundle a set of monochromatic QDs into an ensemble to form a pixel with a minimal lateral area allowing for ultra-high-resolution displays, and with optimal spacing from material interfaces. Consequently, a fabrication technique capable of precise positioning of QD is sought. But the fabrication must also be commercially viable. That bring us to another noticeable shortcoming in the state of the art:

**(e)** A commercially viable fabrication technique.

One of the most commercially viable nano-fabrication techniques for mass production is nano-imprinting. The device is based on a planar design, (well suited for nanoimprinting), and may ultimately be sandwiched between layers of polymers, to produce *flexible* micro-displays. Fabrication of the display over a flat surface and its flexibility would allow the display to be set over a curvature (convex or concave depending on application) eliminating the need for external lenses (which are almost non-existent in small scales due to the diffraction limits). The flexibility of the display and the directionality of its pixel emission would lead to other possibilities and applications.

## 2. Proposed Design and Numerical Results

Figure 1(a) shows the tricolor pixels in a closed packed configuration. A single pixel is marked by a triangle. Figure 1(b) depicts the core elements of proposed pixel design.

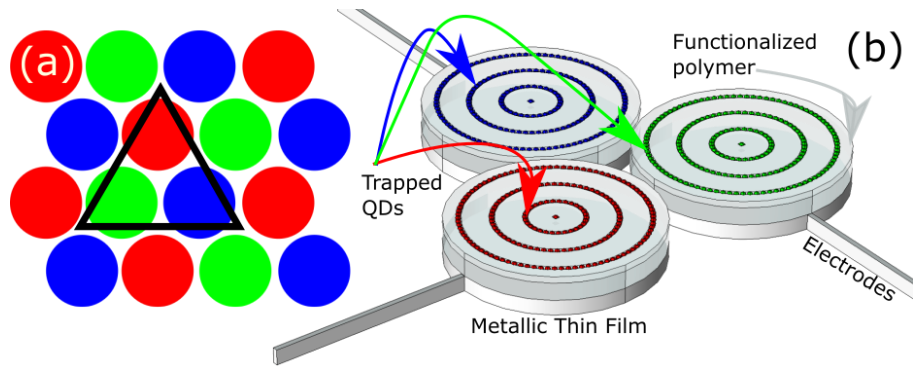

Figure 1: (a) tricolour pixels in a closed packed configuration, (b) Realization of proposed colour pixels and its elements

In my design I have taken advantage of the following effects:

**Beaming effect in optical antenna arrays**. In an array of transmitting optical antennas fixed in a planar and periodic setting, the spacing between the transmitters determines the far-field radiation pattern[7]. The aim here is to maximize the radiation along zero-order mode (m = 0) while suppressing higher order wasteful diffractions.

**Surface Plasmon Coupled Emission (SPCE) and related effects**. Emission of a quantum emitter (such as a QD) is altered by a nearby metallic surface[8-13]. At certain distances, the interaction of induced Surface Plasmon Polaritons SPPs with the QD's dipole moment results in an enhancement to its emission, hence the change in the density of state.

**Coupling of Localized Surface Plasmon Resonance (LSPR) to the radiation.** Metallic thin films (in the shape of disks) under the QD arrays, as depicted in in Figure 1(b), serve multiple purposes. Apart from the already mentioned induced SPCE effect, formation of LSPRs over the disk surface that can decay into freely propagating electromagnetic waves, can play an important role in shaping the radiation pattern of a color element.

2D simulation were carried out with PMMA as polymeric layer having a refractive index 1.5, and underlying silver layer with tabulated refractive index obtained from Palik [14]. To "demonstrate" the impact of SPCE and the beaming effects in action, an ensemble of QDs (eleven in total) were simulated in 2D. The spacing between the QDs denoted with "$P$" in Figure 2(a), was optimized for $\lambda_0 = 645$ nm to produce the highest possible emission along the optical axis. Thickness and the depth of the nano-wells were optimized for a maximal SPCE effect. Figure 2(b) shows electric field intensity $|E|^2$ due to the collective excitations of eleven QDs that takes advantage of both the antenna array and the SPCE concepts. To show the impact of SPCE (or the lack of it), same model was simulated without the silver film, see Figure 2(c) for a drastic reduction in emission. To demonstrate the impact of lateral distribution (vs the vertical staking of QDs), an ensemble of QDs piled vertically (also composed of eleven QDs) were modelled, see the inset of Figure 2(d). The simulated field intensity is depicted in Figure 2(d). To demonstrate the impact of LSPR, a 2D model (similar to Figure 2(a) but with of only two QDs) was simulated. Dimensions of the silver strip (or the disk in 3D) were optimized for $\lambda_0 = 645$ nm and its width (or diameter in 3D) was found to be less than 700 nm, see Figure 2(e) inset. The radiation pattern, $|E|^2$ depicted in Figure 2(e) clearly shows the emission not being wasted by spreading laterally inside the polymer. In summary, the contribution of LSPRs to the overall radiation of the system is another factor that can assist in the reduction of pixel size while maintaining the unidirectional radiation pattern.

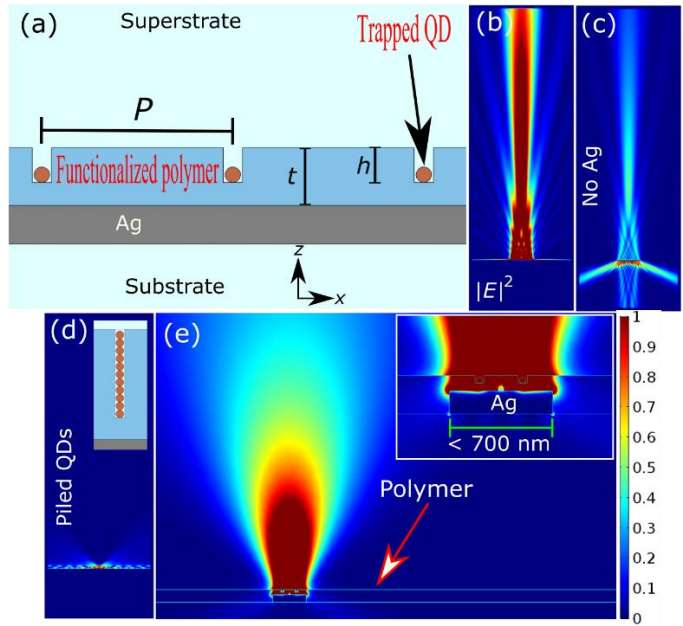

Figure 2: (a) Proposed structure for a single color element and its dimensions. (b) Electric field intensity $E^2$ due to the collective excitations of eleven QDs that takes advantage of both the antenna array and the SPCE concepts. (c) Same as (b) with no silver film (hence no SPCE effect). (d) $|E|^2$ due to the collective excitations of eleven QDs piled on top of each other. Inset: Ensemble of QDs piled vertically. (e) Impact of LSPR. Radiation pattern, $|E|^2$, at $\lambda_0 = 645$ nm of only two QDs positioned above a silver strip with a length less 700 nm

Since I am reporting on a concept, exact dimensions are not crucial in this report. An optimization process would be needed for different choice of material. 3D realization of the concept may be achieved by rotational symmetry about the *z*-axis, where nano-wells take the form of concentric circular corrugations that trap QDs, as depicted in Figure 1(b).

**Localization of electrostatic fields inside the nano-wells**. The other effect sought to be utilized in my design is the electrostatic field distribution in inhomogeneous dielectrics. Figure 3(b) represent the schematic of a single nano-well with a trapped QD, simulated with potential difference applied between the silver plate and the top surfaces of the device. The $E_x^2 + E_y^2$ was calculated over all surfaces (inside and outside the well) with red and blue signifying the maximum and minimum values respectively. Note the high electrostatic field intensity (red in color) localized inside the dielectric nano-well and around the QD. This relatively unknown electrostatic effect would play an important role in the enhancement of quantum yield of trapped QDs, when excited electrically.

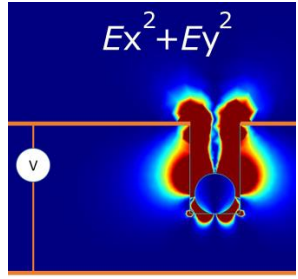

Figure 3: The $E_x^2 + E_y^2$ calculated (inside and outside the well) with red and blue signifying the maximum and minimum values respectively. High electrostatic field intensity localized inside the nano-well and around the QD is noticeable.

**Algorithm for trapping**

Three types of QDs, one for each colour must be identified with each type of QD having a distinct diameter with minimum variation in size. For example, if the average sizes of QDs are denoted by $d_{red}$ and $d_{green}$ for red and green with their respective size variations denoted by $\Delta d_{red}$ and $\Delta d_{green}$, the overlap between $d_{red} - \Delta d_{red}$ and $d_{green} + \Delta d_{green}$ must be minimized. Let us assume $d_{red} > d_{green}$. This is important in eliminating cross-contamination of nano-wells with QDs that are not intended to seed. The idea is to seed the widest nano-wells corresponding to the largest QDs first. This ensures that only nano-wells having width $w_{red} \approx d_{red}$ would participate in the first round of seeding. Once the widest nano-wells are all occupied, seeding of the second widest nano-wells that host the green QDs would follow. To avoid cross-contamination (i.e. to prevent seeding of red QDs in a green or a blue nano-well and green QDs in a blue nano-well) the condition $d_{red} - \Delta d_{red} > d_{green} + \Delta d_{green}$ AND $d_{green} - \Delta d_{green} > d_{blue} + \Delta d_{blue}$ must be satisfied. The width of the wells must also be designed so that $w_{red} > w_{green} > w_{blue}$.

## 3. Funding, acknowledgments, and disclosures

## 4. Summary

The new high-resolution pixel technology proposed here is based on a planar design, (well suited for nanoimprinting), and would ultimately be sandwiched between layers of polymers, to produce flexible micro-displays. The flexibility of the display and the directionality of its pixel emission would lead to many possibilities and applications. It is anticipated that with nanoimprinting fabrication technique, such concepts would be commercially viable. And with its submicron pixel size, it is possible to miniaturize the micro-displays beyond what is available today while improving their resolution.